\documentclass[acmtog]{acmart}

\usepackage{multirow} 
\usepackage{subcaption} 
\usepackage{float} 
\newsavebox{\tempfig} 
\usepackage{amsmath,amsfonts}
\usepackage{algorithmic}

\usepackage{graphicx}
\DeclareGraphicsExtensions{.png,PNG,.eps,.EPS}
\graphicspath{{./images/}}

\usepackage{textcomp}
\usepackage{xcolor}
\def\BibTeX{{\rm B\kern-.05em{\sc i\kern-.025em b}\kern-.08em
    T\kern-.1667em\lower.7ex\hbox{E}\kern-.125emX}}
\usepackage{soul} 

\AtBeginDocument{%
  \providecommand\BibTeX{{%
    \normalfont B\kern-0.5em{\scshape i\kern-0.25em b}\kern-0.8em\TeX}}}

\copyrightyear{2020}
\acmYear{2020}
\setcopyright{acmcopyright}\acmConference[SIN 2020]{13th International Conference
on Security of Information and Networks}{November 4--7, 2020}{Merkez, Turkey}
\acmBooktitle{13th International Conference on Security of Information and Networks
(SIN 2020), November 4--7, 2020, Merkez, Turkey}
\acmPrice{15.00}
\acmDOI{10.1145/3433174.3433605}
\acmISBN{978-1-4503-8751-4/20/09}

\acmSubmissionID{9}

\begin{document}

\title{Visual spoofing in content-based spam detection}

\author{Mark Sokolov}
\email{sokolovm19@students.ecu.edu}
\orcid{1252-266-7423}
\affiliation{%
  \institution{East Carolina University}
  \streetaddress{E 5th Street}
  \city{Greenville}
  \state{NC}
  \postcode{27858}
}

\author{Kehinde Olufowobi}
\affiliation{%
  \institution{East Carolina University}
  \streetaddress{E 5th Street}
  \city{Greenville}
  \state{NC}
  \postcode{27858}}
\email{olufowobik18@students.ecu.edu}

\author{Nic Herndon}
\affiliation{%
  \institution{East Carolina University}
  \streetaddress{E 5th Street}
  \city{Greenville}
  \state{NC}
  \postcode{27858}}
\email{herndonn19@ecu.edu}

\renewcommand{\shortauthors}{Sokolov, Olufowobi, and Herndon}

\begin{abstract}
Although the problem of spam classification seems to be solved, there are still vulnerabilities in the current spam filters that could be easily exploited. We present one such vulnerability, in which one could replace some characters with corresponding characters from a different alphabet. These characters are visually similar, yet have a different Unicode encoding. With this approach spammers can create messages that bypass existing spam filters. Moreover, we show that this approach can be used to avoid plagiarism detection, and in other applications that use natural language processing for automatic analysis of text documents.
\end{abstract}



\maketitle

\section{Introduction}
The electronic mail appeared almost simultaneously with the advent of the Internet. Over the years, it has almost completely replaced traditional mail. Today, most of the communication between people over long distances is done through email. Owing to ease of access, the number of email users today continues to grow rapidly.  The use of email extends beyond just messages in text form; it is also used for sharing other types of data -- images, videos, archive files, etc. \cite{8474778}. As with many technologies intended for the good of humanity, the mass adoption of email as a quick and efficient means of written exchange also enabled the rapid proliferation of spam -- the mass transmission of unwanted messages. Scammers soon began to refine their approach into one that involves the fraudulent use of emails to induce individuals to reveal sensitive or private information, a practice known as phishing \cite{7393036}. The purpose of phishing is to gain access to confidential user data such as usernames and passwords. This is achieved by sending bulk emails on behalf of popular brands, or private messages within various services on behalf of banks or within social networks. The message often contains a direct link to a site that is apparently indistinguishable from the genuine one, or to a site with a redirect. After a user lands on a fake page, scammers try to induce the user to enter their username and password on the counterfeit page using various psychological tricks. Once the user reveals the appropriate credentials, they inadvertently grant access to fraudulent third party.

A spam filter is used in email applications to detect and separate spam from genuine messages based on certain criteria. Different kinds of spam filters are used to achieve this; for example, white list/black list filter, header filter, content-based filter, etc. Black list strategy blocks messages from email addresses, and IP addresses known to be spammers. White list strategy specifies which senders to permit. Header filters check the header of the email for its source, substance of the message and other details contained in the header of the email. Content based filters are, for the most part, used to check the body of messages and decide if the email is spam or not \cite{8527737}. While spammers are unrelenting in developing different approaches for modifying data identifying spam-related words, usually through the expansion of complexities, different machine learning classifiers are also being utilized to combat these tactics.

Different machine learning classifiers have been utilized in the exploration to handle such issues \cite{7743279}. These procedures extract data from prepared datasets and use them to train a classifier \cite{7548905}. The machine learning algorithms that are generally well known in spam classification are na\"ive Bayes \cite{8776800}, support vector machine \cite{8776800}, decision tree \cite{8226246} and random forest \cite{8226246}, as these algorithms are more successful among different techniques at detecting spam messages \cite{8703222}. However, the accuracy of these algorithms is dependent on the data used in training them, as they operate under the assumption that the training data comes from the same distribution as the test data. In practice, this is not always the case.

The main contribution of this work is that it identifies a vulnerability in existing spam filters, which is important because ``learning the vulnerabilities of current classifiers is the only way to fix them'' \cite{10.1145/1014052.1014066}. In particular, we show how substituting letters with their corresponding confusable tricks spam filters into classifying spam emails as ham emails. In addition, we propose methods to address this threat, and enable spam filters to continue to protect people from cybercriminals and losing personal information such as bank account and card data, logins and passwords of Internet services, and so on.

\section{Related Work}
To solve the issues brought about by spam, many spam sifting arrangements were proposed in the ongoing past years. In \cite{6009529}, the authors described a method that used text features that were long established, such as frequency of spam words and HTML tags, as well as some that were new. The novelty of their work was that they introduced language-centric features such as grammar and spell errors, use of function words, presence of verbs and alphanumerics, TF-IDF, and inverse sentence frequency. They evaluated the classifier performance on four benchmark email datasets: CSDMC2010, SpamAssassin, LingSpam, and Enron-Spam. Since the highlights identified with the meaningfulness of email writings are language-independent, the strategy proposed in this paper is conceivably ready to group messages written in any language. The aforementioned features, as well as the traditional ones, are used to generate binary classifiers by five well-known learning algorithms.

In \cite{8474778}, the authors presented different classifiers for detecting spam. They evaluated two main approaches to detect spam: header-based features and content-based features. The classifiers presented in this paper include Support Vector Machine (SVM), na\"ive Bayes (NB) and J48. The dataset utilized in this paper is enron1 from Enron spam. It contains 3762 spam messages and 5172 ham messages. To assess their effectiveness they compute accuracy, precision and recall. They found that SVM is the best classifier as far as accuracy and False Positive Rate are concerned.  

In \cite{8282698}, the authors evaluated SVM classifiers with different values for the C parameter, given that SVM is one of the best algorithms when it comes to text analysis and prediction. Their observation was that for high values of C the model overfits, and for low values of C the model underfits, thus highlighting the importance of choosing the appropriate value for the C parameter.

In \cite{8527737}, the authors proposed a weighted SVM method for spam filtering. This method used weight variables obtained by KFCM algorithm. They evaluated this method with emails from the UCI Repository SMS Spam base dataset, and compared with SVM and Improvised WSVM. Based on their analysis, Improvised WSVM produced lower misclassification rates that SVM.

In \cite{8455990}, the authors proposed a method that used the na\"ive Bayes algorithm with word features in which symbols within words are replaced by the letter that most likely substitutes that symbol. With this change, their method increased the classification accuracy by over two hundred percent over Spamassassin. They evaluated this method using the Ling-Spam corpus, a dataset that best emulates genuine circumstances.

Dalvi et al. \cite{10.1145/1014052.1014066} formalized the problem of adversarial classification -- in the context of spam filtering, this is a continuous competition between the spam filter designers, and an adversary that attempts to make the classifiers used in spam filters, to mark spam messages as harmless. They analyzed the optimal strategies of the adversary and of the spam filter classifiers, using the naive Bayes as an example. They showed that taking into account the adversary’s optimal feature-changing strategy, consistently outperforms the standard classifier, sometimes by a large margin. They do point out though, that complete automation will never be possible, and spam filters have to be updated to address new threats from the adversary.

Lowd and Meek \cite{10.1145/1081870.1081950} extended the analysis of adversarial learning, and pointed out that the assumption that an attacker has perfect knowledge of the classifier, under which the Dalvi et al. model operates, is unrealistic. They proposed instead that an adversary has to reverse engineer the learning problem, and learn sufficient information about a classifier to construct effective adversarial attacks. Moreover, they believed that \textit{learning the vulnerabilities of current classifiers is the only way to fix them in the future}. Towards this goal, they proposed a theoretical framework for studying adversaries and classifiers, which can be employed to determine whether an adversary can efficiently learn enough about a classifier to minimize the cost of defeating it.

Barreno et al. \cite{10.1145/1128817.1128824} approached the adversarial learning with the goal of answering the question, "Can machine learning be secure?" They classified the different types of attacks and the corresponding defenses, and proposed an analytical model to predict the lower bound on the adversary’s capability when the adversary has complete control of the learner’s training.

\section{Experimental design}
\begin{figure}[tbhp]
    \centering
    \includegraphics[width=0.48\textwidth]{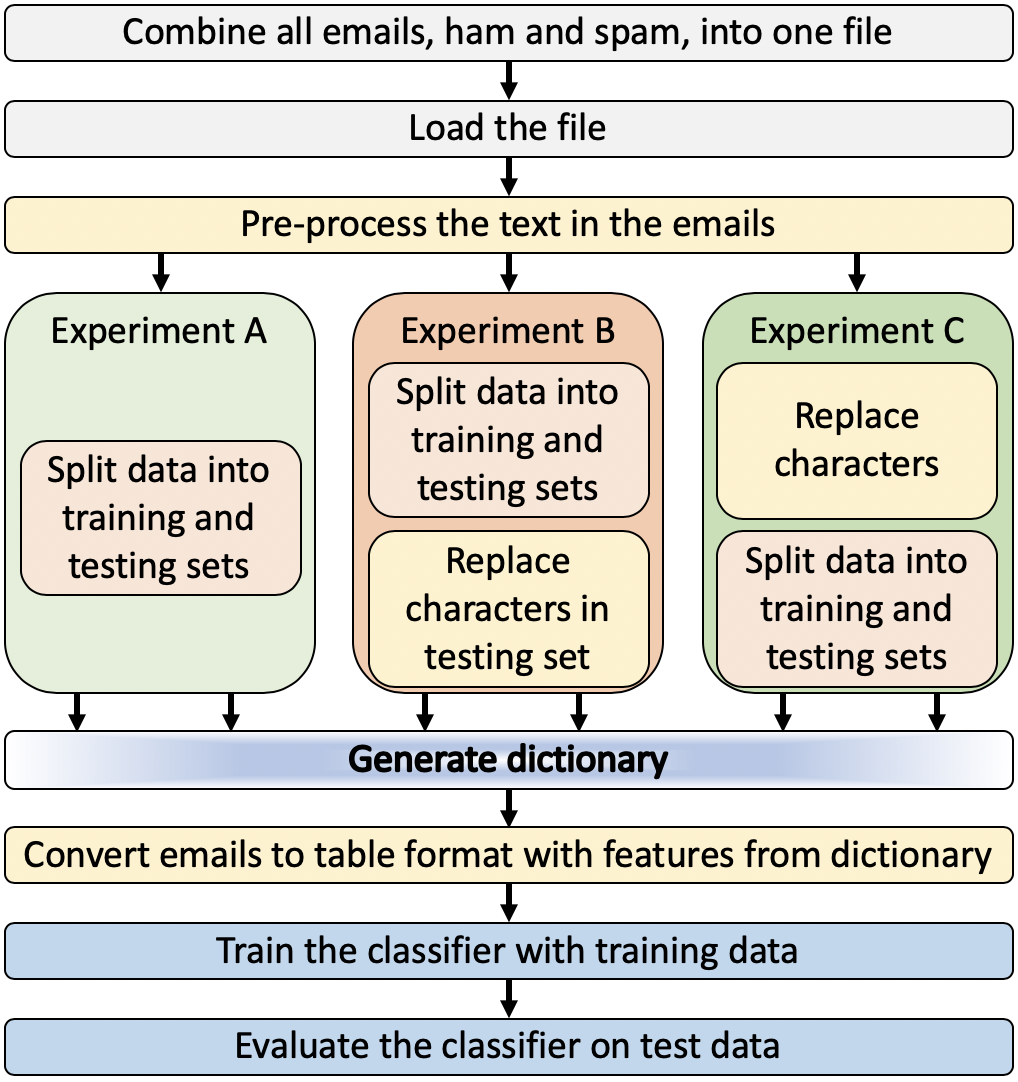}
    \caption{Experimental design: The emails from the Enron1 dataset are combined into one file, with each email having its associated label, ham or spam. This file is loaded in memory and preprocessed prior to analysis, by replacing URLs, emails, phone numbers, and numbers with their corresponding placeholders, followed by removing stop words and punctuation symbols, and then stemming the words. Before generating the dictionary from these processed emails, in Experiments B and C, some of the characters are replaced with their corresponding confusables from the Cyrillic alphabet. The emails are then converted to a table format required by the machine learning algorithms, with each email represented by a vector indicating the presence or absence of the words from the dictionary in the body of the email. With this representation the emails from the training set are used to train the classifiers, which are then evaluated with the emails from the testing set.}
    \label{fig:experimental_design}
\end{figure}

The goal of our experimental design is to assess the effect of confusables on the performance of typical content-based spam detection machine learning models. To this end, three different experiments were conducted. Experiment A is the control experiment involving entirely unmodified datasets -- the default encoding of characters in both the training and testing sets is preserved. In Experiment B, the encoding of the training set is preserved whereas the encoding of certain characters in the testing set is switched from the default Latin alphabet to their corresponding confusables from the Cyrillic alphabet. In Experiment C, confusables are introduced to both the training and testing sets so that each model is trained and evaluated with data from a single, mixed-script that contains confusables. The steps of each experiment are shown in Fig.~\ref{fig:experimental_design}.

\subsection{Data Source and Format}

The experiments were performed using the emails from the Enron1 dataset, a preprocessed subset of the publicly available Enron Corpus \cite{metsis2006spam}. Enron1 consists of 1500 spam emails and 3672  legitimate or “ham” emails stored as plain text documents in separate files. The dataset and relevant metadata is available online at \url{http://www2.aueb.gr/users/ion/data/enron-spam/}. The experiments were run entirely in a Python environment using the scikit-learn machine learning library \cite{scikit-learn}, the Natural Language Toolkit (NLTK) \cite{bird2009natural}, and using pandas library \cite{mckinney-proc-scipy-2010} for data manipulation. We reformatted the Enron1 dataset to suit this setup by pooling the entire set of email messages in a single comma-separated values file such that each row is a full representation of an individual email’s information consisting of a ``spam'' or ``ham'' label, comma-separated from the email text. The reformatted file was parsed and converted to a pandas dataframe with no headers. The resulting dataframe is a simple framework that consists of two columns only: the email instance and the corresponding label, with the label appearing before the email instance. Each email message is a single space-delimited stream of words. 

\subsection{Text Preprocessing}
We preprocessed each email by replacing any email addresses, URLs, currency symbols, and numbers with suitable string placeholders that identifies the original token. Each token that does not start with with letters, digits, or spaces was replaced with a space, and multiple spaces were trimmed to a single space. Empty lines and stop words were also removed. Stemming was done using the Porter stemming algorithm \cite{van1980new} from NLTK. Using the scikit-learn dataset splitting utility, we split our dataset into random training and testing subsets such that 80 percent of the sample is devoted to model training and the other 20 percent to testing. 

\subsection{Feature Extraction}
Word tokenization and feature creation for our word dictionary was carried out using the NLTK word tokenization and frequency distribution utilities. The label encoder utility in scikit-learn was used to generate features from the words in emails.

\subsection{Experiment A -- Training and Test Data without Confusables}
Each email text was preserved in its original form -- i.e., no characters are replaced. We trained and evaluated our set of classifiers using data from this distribution only. 

\subsection{Experiment B -- Training Data without Confusables, and Test Data with Confusables}
We modified our testing data by introducing corresponding Cyrillic letters in place of the Latin letters `a', `e', `k', `o', `p', `c', and `y'. As desired, this resulted in single and mixed-script confusables in our testing dataset. The intent here was to simulate a visual spoofing effect in the email messages that our testing set consists of. With the original character encoding in our training dataset still preserved, we trained all classifiers using data from the same distribution as in experiment A. However, model evaluation is done using data effectively from a different distribution than the one used in experiment A.

\subsection{Experiment C -- Training and Test Data with Confusables}
We modified both our training and testing datasets to replicate visual spoofing in the entire distribution used to develop our model, using the set of confusables introduced in experiment B. Training and evaluation was performed using data from this modified distribution only. As a result, unlike experiments A or B, each of our models would simulate spam filters designed to classify emails that contain visual spoofing.

\subsection{Models and Model Evaluation Metrics}
We evaluated these three scenarios with four machine learning algorithms frequently used for spam detection: decision tree, random forest, na\"ive Bayes, and support vector machine. The goal was not to compare these algorithms, but rather to show that regardless of the algorithm used, this method leads to similar results. Each of these classifiers was evaluated using accuracy, precision, recall, F1 score, and confusion matrices.

\subsection{Production testing}
\begin{figure*}[tbhp]
    \centering
    \begin{subfigure}[t]{0.5\textwidth}
        \centering
        \includegraphics[width=0.9\textwidth]{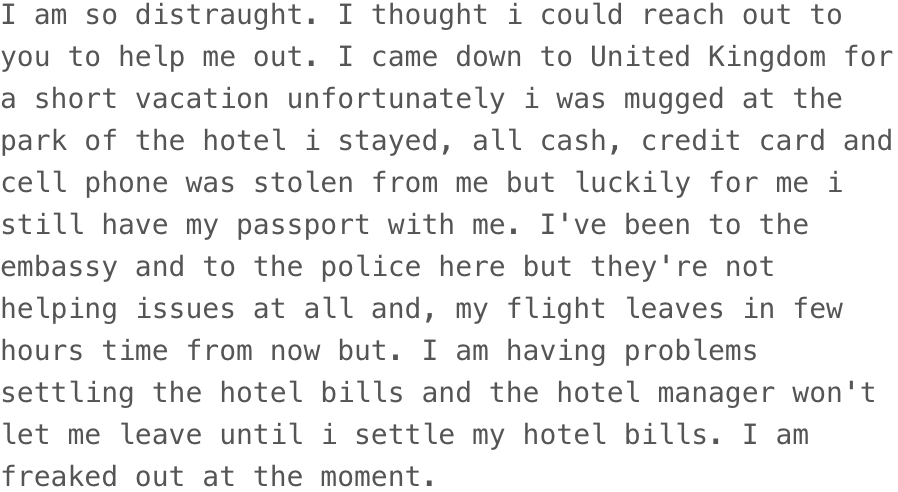}
        \caption{Text unchanged}\label{fig:Production testing with unchanged text}
    \end{subfigure}%
    ~ 
    \begin{subfigure}[t]{0.5\textwidth}
        \centering
        \includegraphics[width=0.9\textwidth]{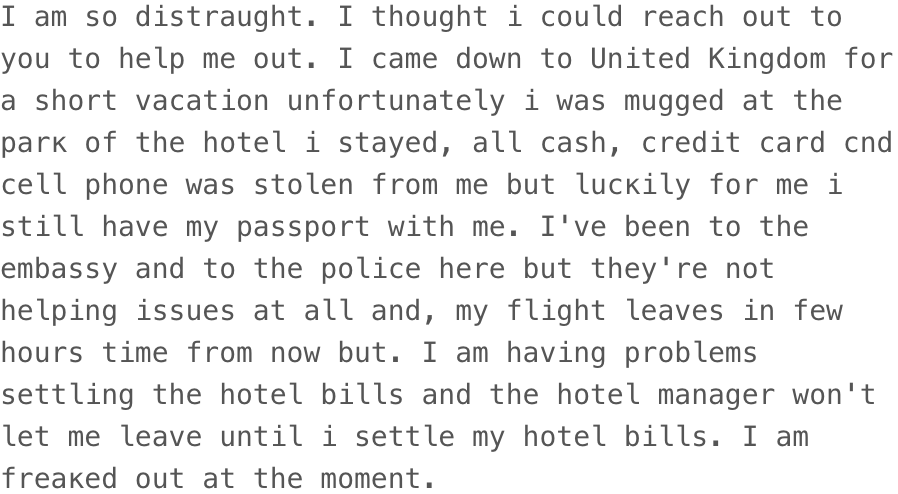}
        \caption{Text changed}\label{fig:Production testing with modified text}
    \end{subfigure}

    \begin{subfigure}[t]{0.5\textwidth}
        \centering

        \savebox{\tempfig}{\includegraphics[width=0.9\textwidth]{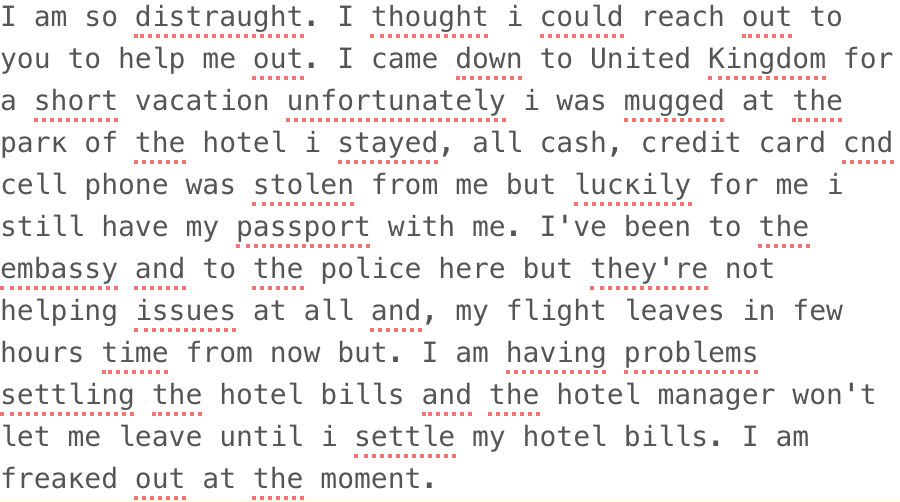}}
        \raisebox{\dimexpr\ht\tempfig-\height}{
        \begin{tabular}{c | c}
          \textbf{Character} & \textbf{Replacement} \\ \hline
          a                  & U+0430               \\
          e                  & U+0435               \\
          k                  & U+043A               \\
          o                  & U+043E               \\
          p                  & U+0440               \\
          c                  & U+0441               \\
          y                  & U+0443              
        \end{tabular}
        }
        \caption{Characters replaced in text}
    \end{subfigure}%
    ~ 
    \begin{subfigure}[t]{0.5\textwidth}
        \centering
        \includegraphics[width=0.9\textwidth]{text_changed_spell_check.png}
        \caption{Spell checker highlights ``misspelled'' words}
    \end{subfigure}

    \begin{subfigure}[t]{0.5\textwidth}
        \centering
        \includegraphics[width=0.9\textwidth]{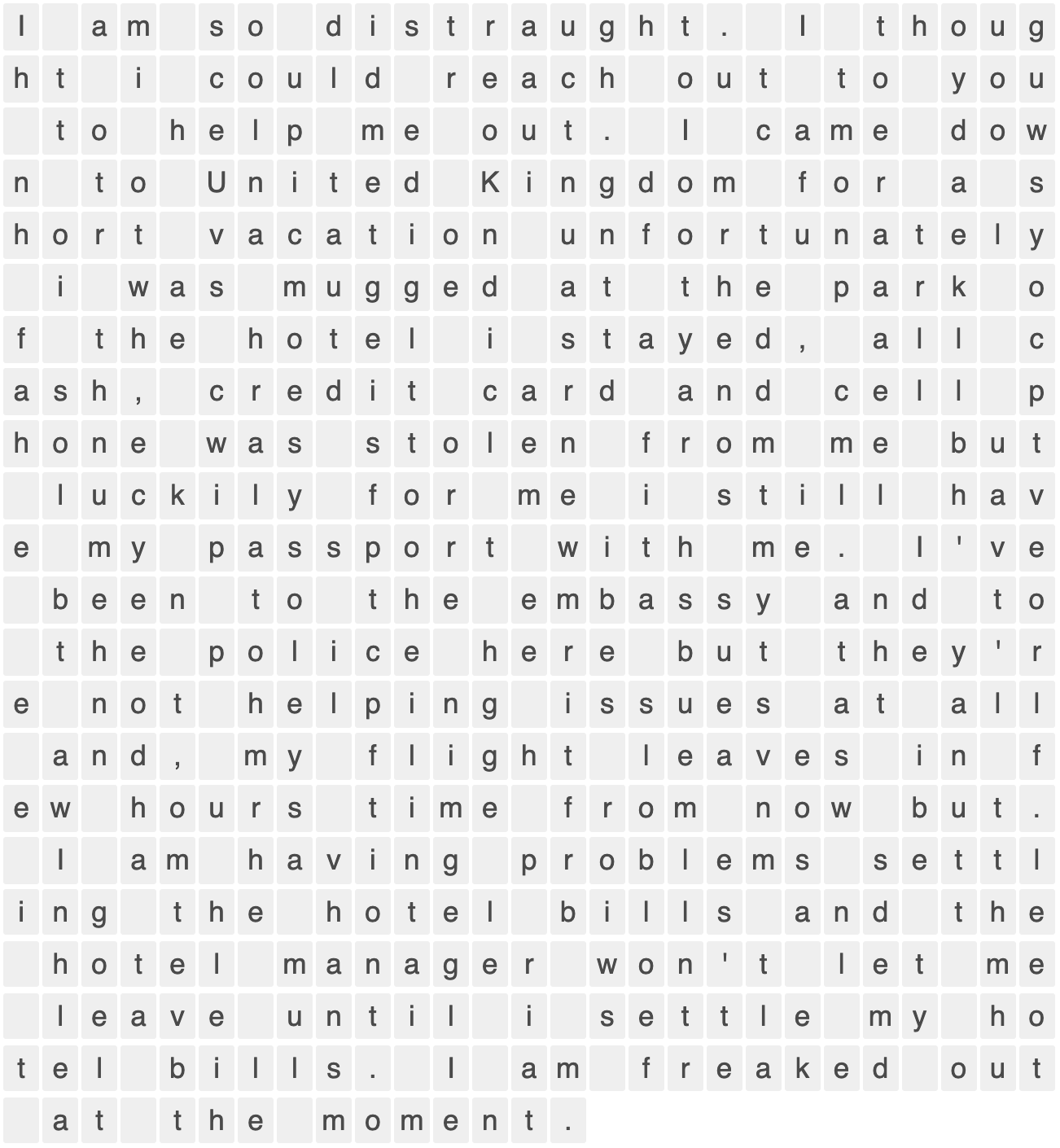}
        \caption{Unchanged text shown in TextMagic}
    \end{subfigure}%
    ~ 
    \begin{subfigure}[t]{0.5\textwidth}
        \centering
        \includegraphics[width=0.9\textwidth]{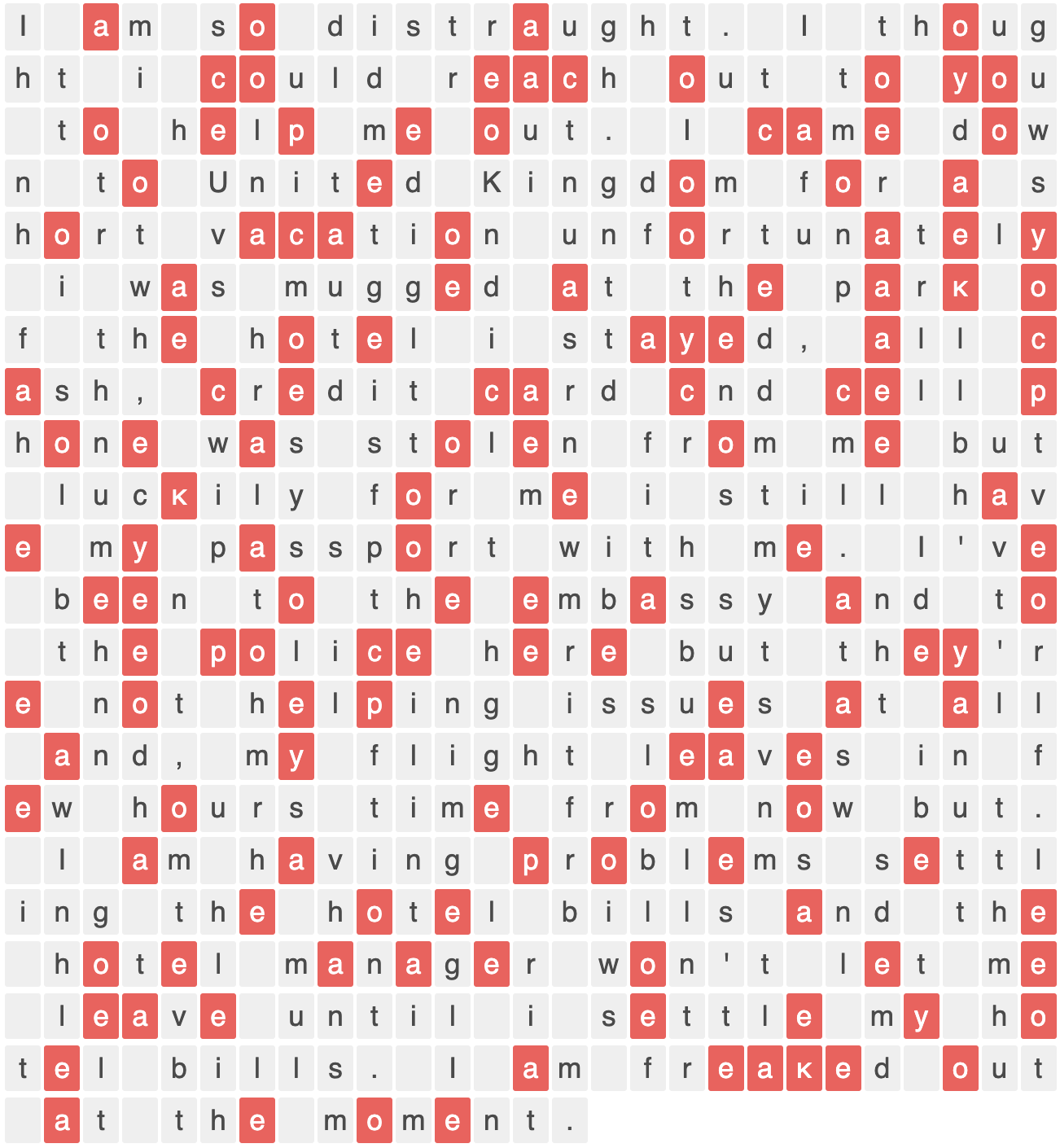}
        \caption{Changed text shown in TextMagic}
    \end{subfigure}
    \caption{An example of text emailed with two different outcomes. The text in sub-figure (a) uses the Latin alphabet, and was detected by the spam filter. The text in sub-figure (b) has some characters replaced with their ``visual-equivalent'' from the Cyrillic alphabet, and was not detected by the spam filter. The characters replaced, and their replacements are shown in sub-figure (c). Some of the words in the modified text cause the spell checker to highlight them, as shown in sub-figure (d). The initial text and the modified text are shown side-by-side in \href{https://www.textmagic.com/free-tools/unicode-detector}{TextMagic}, an online application that shows non-ASCII characters with red background, sub-figures (e) and (f), respectively.}
\end{figure*}

We also tested this method with a production email. We first sent an email containing a lot of keywords frequently encountered in spam emails (Fig.~\ref{fig:Production testing with unchanged text}), and this email was flagged as spam. Then we sent the same email, with some of the characters replaced by their ``visually equivalent'' characters from Cyrillic alphabet, and this email was delivered to the Inbox (Fig.~\ref{fig:Production testing with modified text}). This suggests that this method can currently bypass existing spam filters.

\section{Results and Discussion}\label{Results and Discussion}

\begin{figure*}[tbhp]
\centering
\begin{subfigure}[t]{0.30\textwidth} 
    \centering
    \makebox[10pt][r]{\makebox[20pt]{\raisebox{60pt}{\rotatebox[origin=c]{90}{Accuracy}}}}%
    \includegraphics[width=\textwidth]{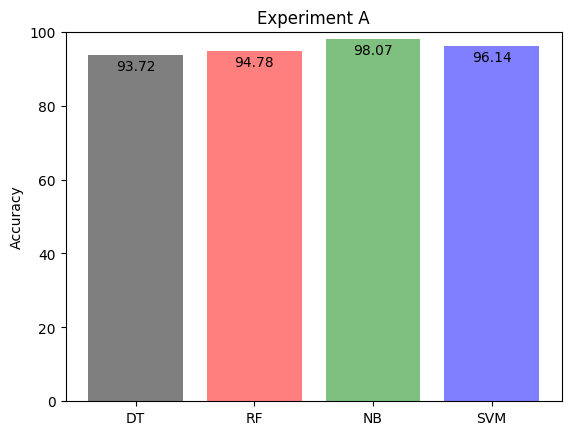}
    
    \vspace{2mm}
    
    \makebox[10pt][r]{\makebox[20pt]{\raisebox{60pt}{\rotatebox[origin=c]{90}{Precision}}}}%
    \includegraphics[width=\textwidth]{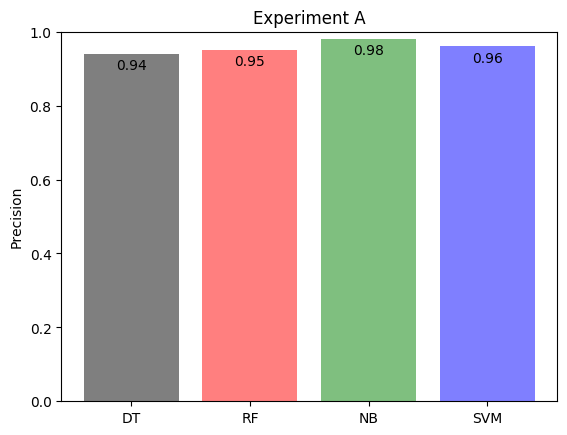}
    
    \vspace{2mm}
    
    \makebox[10pt][r]{\makebox[20pt]{\raisebox{60pt}{\rotatebox[origin=c]{90}{Recall}}}}%
    \includegraphics[width=\textwidth]{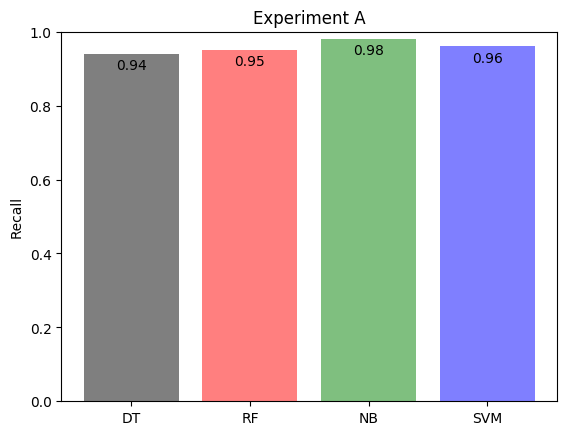}
    
    \vspace{2mm}
    
    \makebox[10pt][r]{\makebox[20pt]{\raisebox{60pt}{\rotatebox[origin=c]{90}{F1 score}}}}%
    \includegraphics[width=\textwidth]{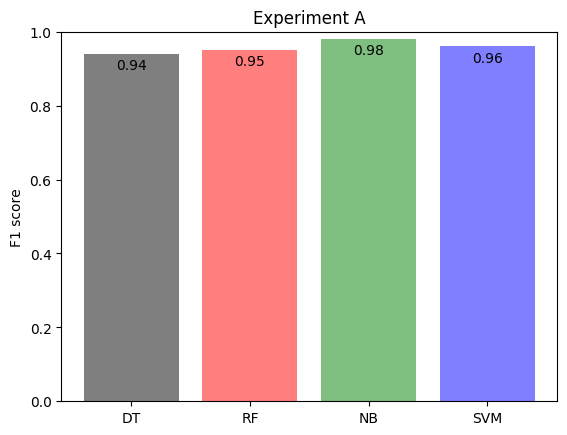}
    \caption{Experiment A}\label{Metrics: Experiment A}
\end{subfigure}
\hspace{1em}
\begin{subfigure}[t]{0.30\textwidth} 
    \centering
    \includegraphics[width=\textwidth]{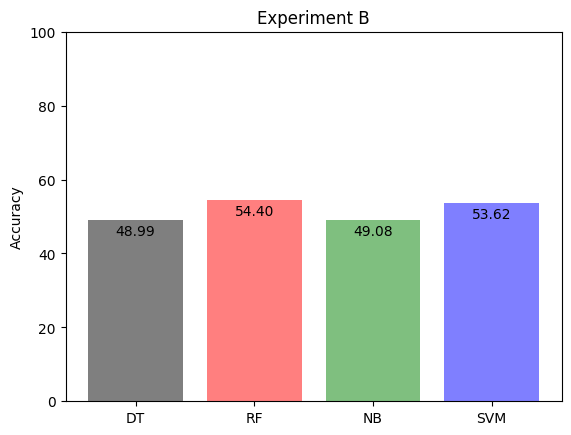}
    
    \vspace{2mm}
    
    \includegraphics[width=\textwidth]{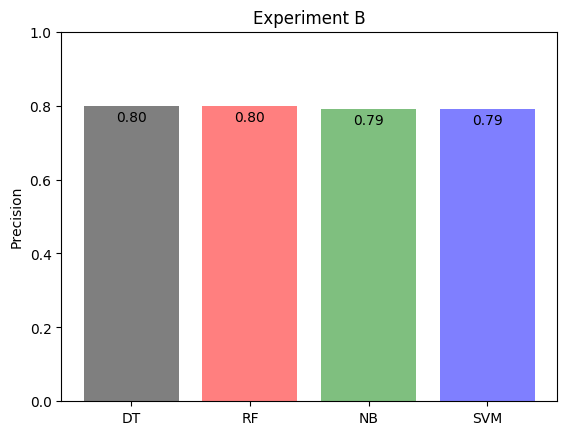}
    
    \vspace{2mm}
    
    \includegraphics[width=\textwidth]{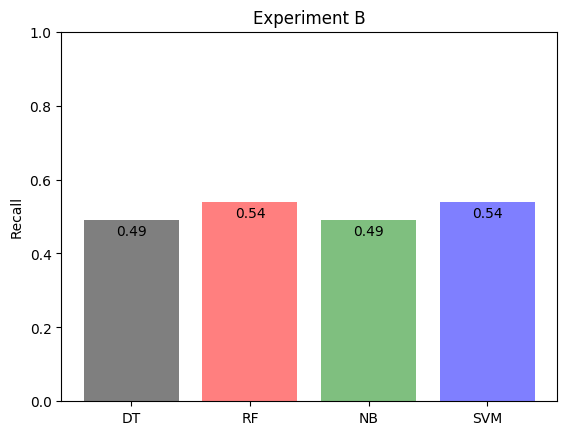}
    
    \vspace{2mm}
    
    \includegraphics[width=\textwidth]{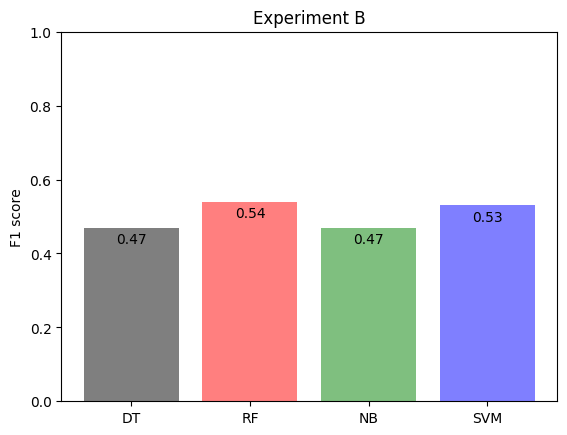}
    \caption{Experiment B}\label{Metrics: Experiment B}
\end{subfigure}
\hspace{1em}
\begin{subfigure}[t]{0.30\textwidth} 
    \centering
    \includegraphics[width=\textwidth]{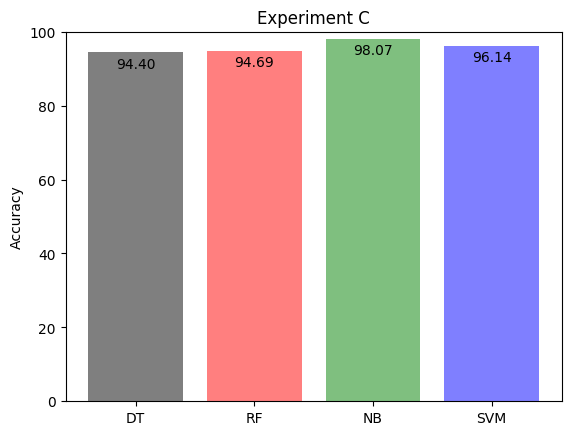}
    
    \vspace{2mm}
    
    \includegraphics[width=\textwidth]{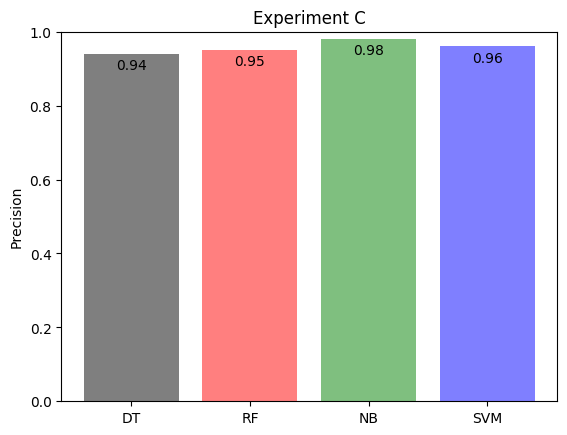}
    
    \vspace{2mm}
    
    \includegraphics[width=\textwidth]{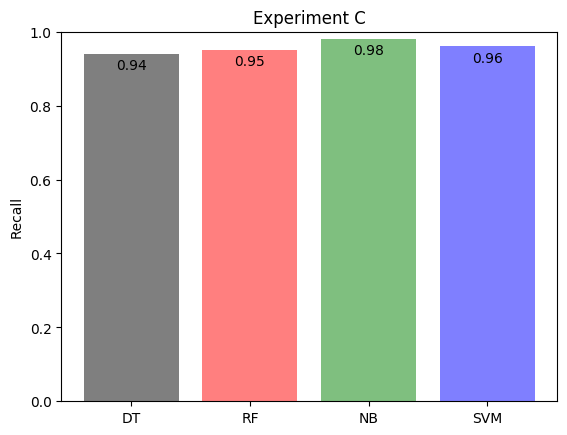}
    
    \vspace{2mm}
    
    \includegraphics[width=\textwidth]{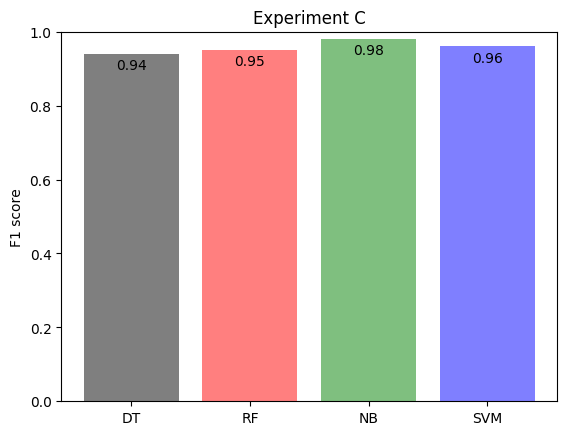}
    \caption{Experiment C}\label{Metrics: Experiment C}
\end{subfigure}
\caption{Evaluation metrics for Experiment A (no characters replaced), Experiment B (characters replaced in the test set), and Experiment C (characters replaced in train and test sets) from the following machine algorithms used: decision tree (DT), random forest (RF), na\"ive Bayes (NB), and support vector machine (SVM). Notice that when replacing characters only in the test set, i.e., Experiment B, the classifiers misclassify most emails than in the other two experiments, and all metrics have lowest values for this experiment.}
\end{figure*}

Experiment A simulates an existing spam filter, by assuming the same distribution for training and testing data. In this experiment we did not replace any characters, and all classifiers correctly identified most of the emails as either ham or spam, as shown in Fig.~\ref{Confusion matrices: Experiment A}. All evaluation metrics used -- accuracy, precision, recall, and F1 score -- had values close to 100\% for all classifiers, as shown in Fig.~\ref{Metrics: Experiment A}.

\begin{figure*}[tbhp]
\centering
\begin{subfigure}[t]{0.30\textwidth} 
    \centering
    \makebox[0pt][r]{\makebox[30pt]{\rotatebox[origin=c]{90}{DT}}}%
    \begin{tabular}{|l|c|c|c|}
      \hline
      & \multicolumn{3}{c|}{predicted} \\ \hline
      \multirow{3}{*}{\rotatebox[origin=c]{90}{actual}} & & ham & spam\\ \cline{2-4}
      & ham & 701 & 30\\ \cline{2-4}
      & spam & 35  & 269\\ \hline
    \end{tabular}
    
    \vspace{2mm}
    
    \makebox[0pt][r]{\makebox[30pt]{\rotatebox[origin=c]{90}{RF}}}%
    \begin{tabular}{|l|c|c|c|}
      \hline
      & \multicolumn{3}{c|}{predicted} \\ \hline
      \multirow{3}{*}{\rotatebox[origin=c]{90}{actual}} & & ham & spam\\ \cline{2-4}
      & ham & 712 & 19\\ \cline{2-4}
      & spam & 35  & 269\\ \hline
    \end{tabular}
    
    \vspace{2mm}
    
    \makebox[0pt][r]{\makebox[30pt]{\rotatebox[origin=c]{90}{NB}}}%
    \begin{tabular}{|l|c|c|c|}
      \hline
      & \multicolumn{3}{c|}{predicted} \\ \hline
      \multirow{3}{*}{\rotatebox[origin=c]{90}{actual}} & & ham & spam\\ \cline{2-4}
      & ham & 725 & 6\\ \cline{2-4}
      & spam & 14  & 290\\ \hline
    \end{tabular}
    
    \vspace{2mm}
    
    \makebox[0pt][r]{\makebox[30pt]{\rotatebox[origin=c]{90}{SVM}}}%
    \begin{tabular}{|l|c|c|c|}
      \hline
      & \multicolumn{3}{c|}{predicted} \\ \hline
      \multirow{3}{*}{\rotatebox[origin=c]{90}{actual}} & & ham & spam\\ \cline{2-4}
      & ham & 717 & 14\\ \cline{2-4}
      & spam & 26  & 278\\ \hline
    \end{tabular}
    \caption{Experiment A}\label{Confusion matrices: Experiment A}
\end{subfigure}
\hspace{2mm}
\begin{subfigure}[t]{0.30\textwidth} 
    \centering
    \begin{tabular}{|l|c|c|c|}
      \hline
      & \multicolumn{3}{c|}{predicted} \\ \hline
      \multirow{3}{*}{\rotatebox[origin=c]{90}{actual}} & & ham & spam\\ \cline{2-4}
      & ham & 207 & 524\\ \cline{2-4}
      & spam & 4  & 300\\ \hline
    \end{tabular}
    
    \vspace{2mm}
    
    \begin{tabular}{|l|c|c|c|}
      \hline
      & \multicolumn{3}{c|}{predicted} \\ \hline
      \multirow{3}{*}{\rotatebox[origin=c]{90}{actual}} & & ham & spam\\ \cline{2-4}
      & ham & 267 & 464\\ \cline{2-4}
      & spam & 8  & 296\\ \hline
    \end{tabular}
    
    \vspace{2mm}
    
    \begin{tabular}{|l|c|c|c|}
      \hline
      & \multicolumn{3}{c|}{predicted} \\ \hline
      \multirow{3}{*}{\rotatebox[origin=c]{90}{actual}} & & ham & spam\\ \cline{2-4}
      & ham & 211 & 520\\ \cline{2-4}
      & spam & 7  & 297\\ \hline
    \end{tabular}
    
    \vspace{2mm}
    
    \begin{tabular}{|l|c|c|c|}
      \hline
      & \multicolumn{3}{c|}{predicted} \\ \hline
      \multirow{3}{*}{\rotatebox[origin=c]{90}{actual}} & & ham & spam\\ \cline{2-4}
      & ham & 262 & 469\\ \cline{2-4}
      & spam & 11  & 293\\ \hline
    \end{tabular}
    \caption{Experiment B}\label{Confusion matrices: Experiment B}
\end{subfigure}
\hspace{2mm}
\begin{subfigure}[t]{0.30\textwidth} 
    \centering
    \begin{tabular}{|l|c|c|c|}
      \hline
      & \multicolumn{3}{c|}{predicted} \\ \hline
      \multirow{3}{*}{\rotatebox[origin=c]{90}{actual}} & & ham & spam\\ \cline{2-4}
      & ham & 704 & 27\\ \cline{2-4}
      & spam & 31  & 273\\ \hline
    \end{tabular}
    
    \vspace{2mm}
    
    \begin{tabular}{|l|c|c|c|}
      \hline
      & \multicolumn{3}{c|}{predicted} \\ \hline
      \multirow{3}{*}{\rotatebox[origin=c]{90}{actual}} & & ham & spam\\ \cline{2-4}
      & ham & 711 & 20\\ \cline{2-4}
      & spam & 35  & 269\\ \hline
    \end{tabular}
    
    \vspace{2mm}
    
    \begin{tabular}{|l|c|c|c|}
      \hline
      & \multicolumn{3}{c|}{predicted} \\ \hline
      \multirow{3}{*}{\rotatebox[origin=c]{90}{actual}} & & ham & spam\\ \cline{2-4}
      & ham & 725 & 6\\ \cline{2-4}
      & spam & 14  & 290\\ \hline
    \end{tabular}
    
    \vspace{2mm}
    
    \begin{tabular}{|l|c|c|c|}
      \hline
      & \multicolumn{3}{c|}{predicted} \\ \hline
      \multirow{3}{*}{\rotatebox[origin=c]{90}{actual}} & & ham & spam\\ \cline{2-4}
      & ham & 717 & 14\\ \cline{2-4}
      & spam & 26  & 278\\ \hline
    \end{tabular}
    \caption{Experiment C}\label{Confusion matrices: Experiment C}
\end{subfigure}

\caption{Confusion matrices for Experiment A (no characters replaced), Experiment B (characters replaced in the test set), and Experiment C (characters replaced in train and test sets) from the following machine algorithms used: decision tree (DT), random forest (RF), na\"ive Bayes (NB), and support vector machine (SVM). Notice that when replacing characters only in the test set, i.e., Experiment B, most of the ham emails get misclassified as spam.}
\end{figure*}

Experiment B simulates an approach that can be used to mislead a spam filter, by changing the distribution of the test data from the distribution of the training data through the use of characters assumed either not present, or rarely present in the training data. With this change, most of the emails are mislabeled, as shown in Fig.~\ref{Confusion matrices: Experiment B}, and accuracy, recall, and F1 score decrease by about half, whereas the precision decreases by about 20\%, as shown in Fig.~\ref{Metrics: Experiment B}.

Experiment C simulates one way a spam filter can adapt to detect emails containing characters replaced with their ``visually equivalent'' counterparts. With this approach, the training and testing data are assumed to be drawn from the same distribution, i.e., some characters are replaced in both data sets. With this modification, all classifiers correctly identified most of the emails as either ham or spam, as shown in Fig.~\ref{Confusion matrices: Experiment C}. In addition, all evaluation metrics used had values close to 100\% for all classifiers, as shown in Fig.~\ref{Metrics: Experiment C}.

Another way to address this situation is to detect the language used in the email, and then replace all characters to the alphabet used with that language. After the characters from a different alphabet are replaced, the training and testing data can be assumed to be drawn from the same distribution, and then the experiment would be similar to experiment A.

\begin{figure*}[tbhp]
\centering
\begin{subfigure}[t]{0.45\textwidth} 
    \centering
\includegraphics[width=0.78\textwidth]{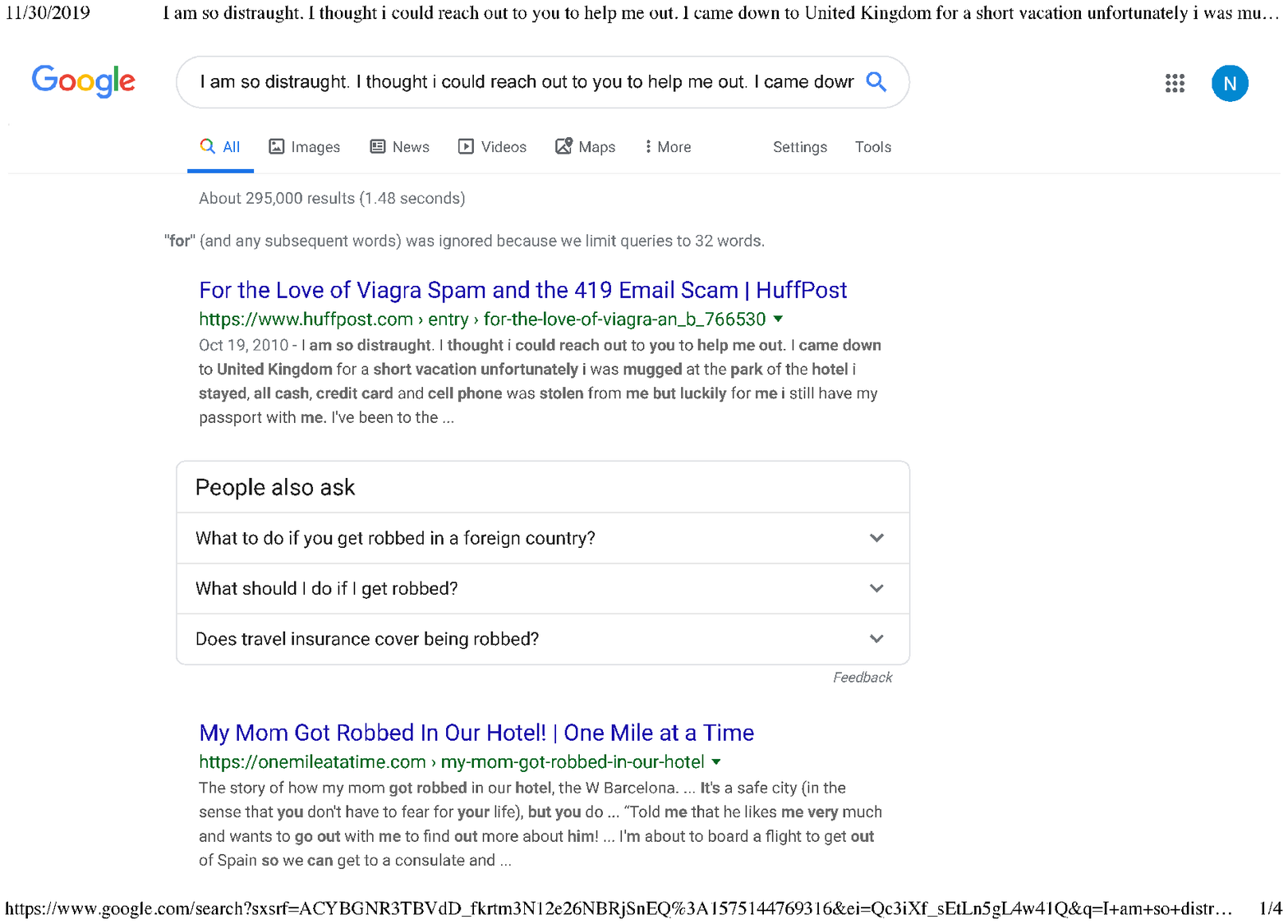}
    \caption{Results returned by Google Search when the input text contained only characters from the Latin alphabet}\label{fig:Google search: Experiment A}
\end{subfigure}
\hspace{2mm}
\begin{subfigure}[t]{0.45\textwidth} 
    \centering
    \includegraphics[width=0.78\textwidth]{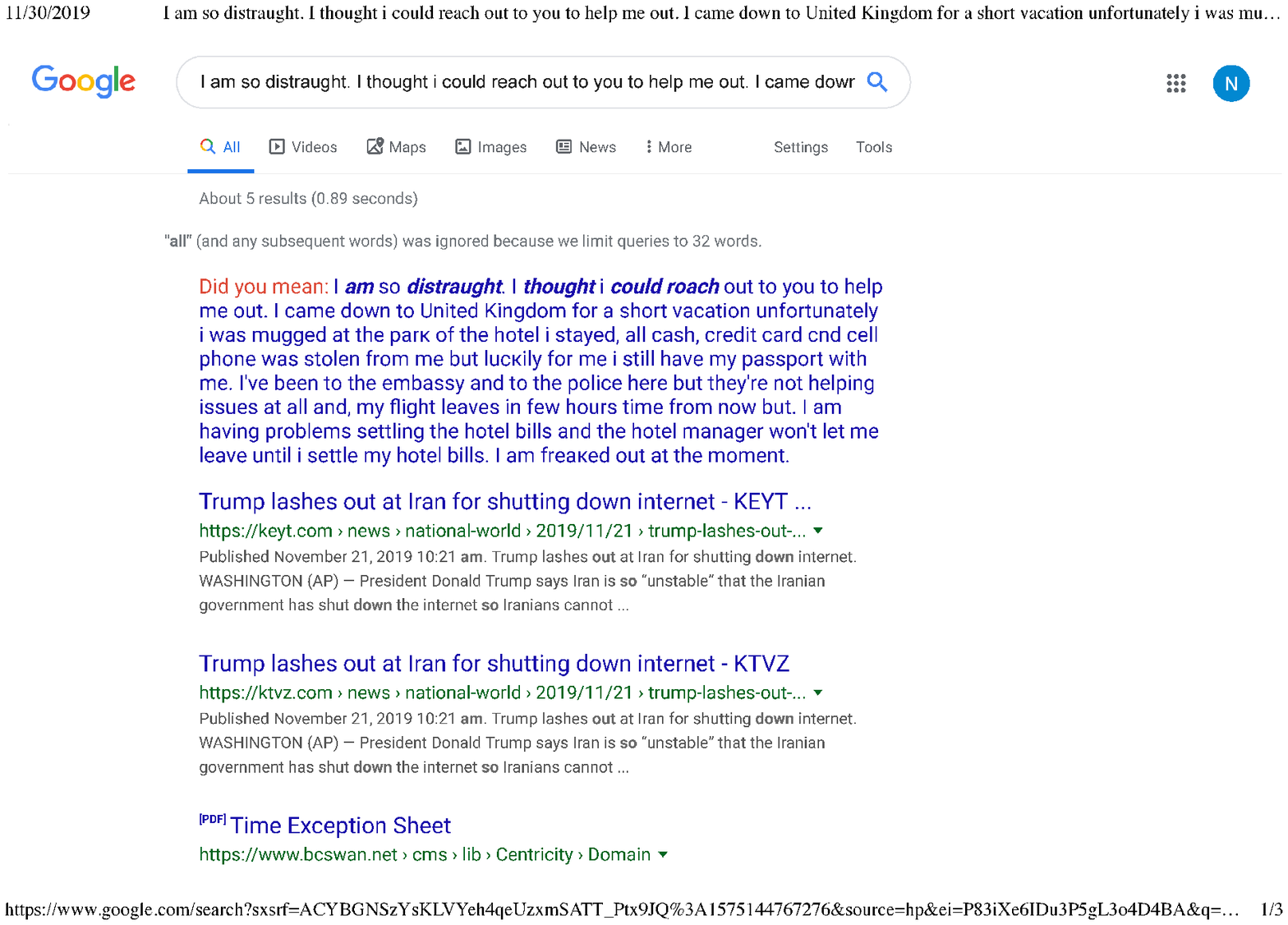}
    \caption{Results returned by Google Search when the input text contained characters from both the Latin and the Cyrillic alphabets}\label{fig:Google search: Experiment B}
\end{subfigure}

\caption{Google Search returns the document containing the searched text as the top result when the characters are not changed, yet it does not find the same document when some characters are changed.}\label{fig:Google search}
\end{figure*}

\begin{figure*}[tbhp]
  \centering
  \includegraphics[width=\textwidth]{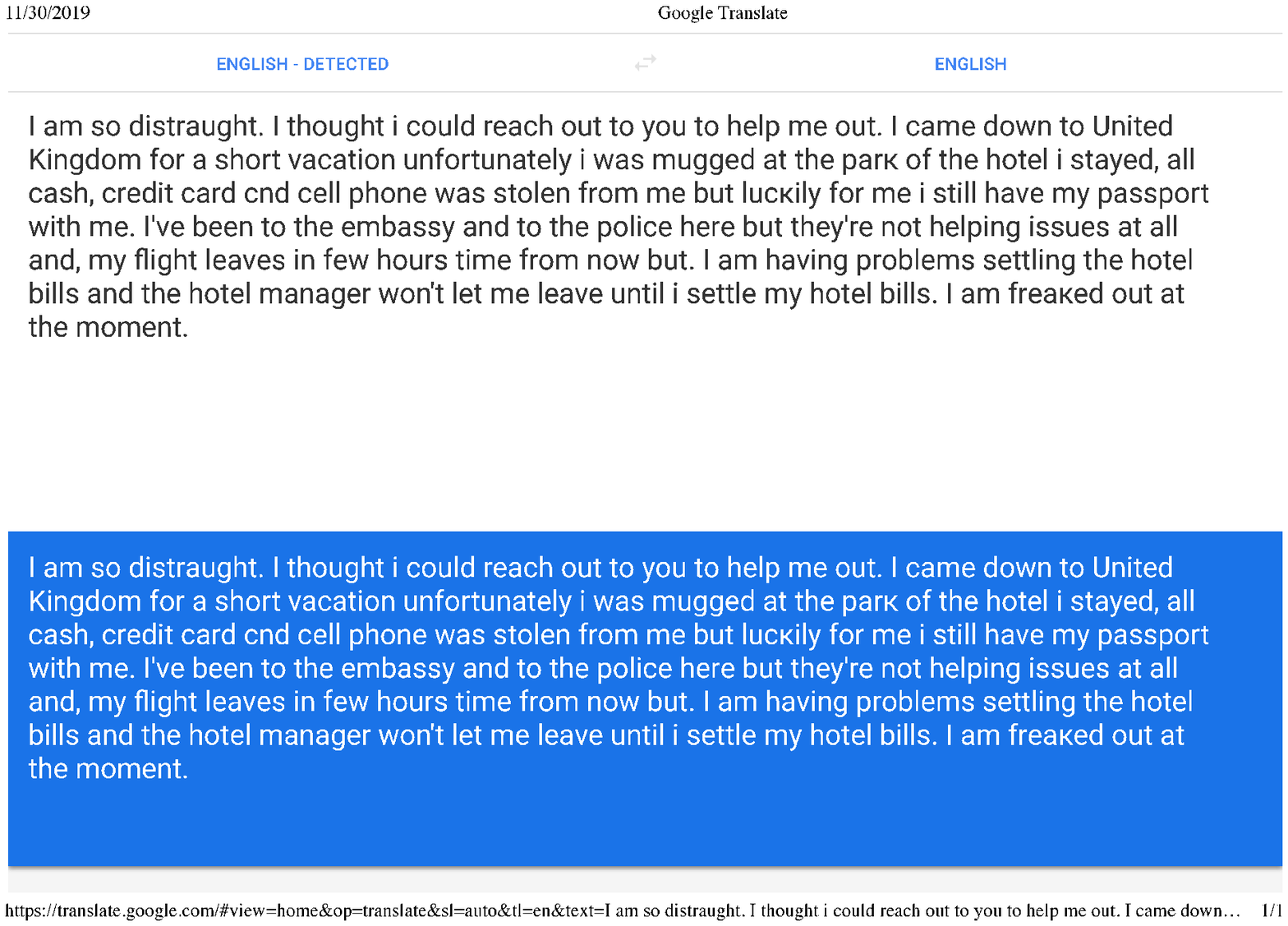}
  \caption{Google Translate correctly identified the language even with confusables, yet it left those characters unchanged in the ``translated'' text.}\label{fig:Google translate}
\end{figure*}

We also evaluated how replacing some characters affect other applications used with text classification:
\begin{itemize}
  \item For web search, replacing some characters with their confusables caused the search engine to not find the document, as shown in Fig.~\ref{fig:Google search}. This suggests that applications used to detect plagiarism could also be affected.
  \item For text translation, Google was able to correctly identify the language even with confusables used in text, yet it left those characters unchanged in the ``translated'' text (Fig.~\ref{fig:Google translate}).
\end{itemize}

\begin{figure*}[htbp]
\centering
\begin{subfigure}[t]{0.48\textwidth} 
    \centering
    \includegraphics[width=\textwidth]{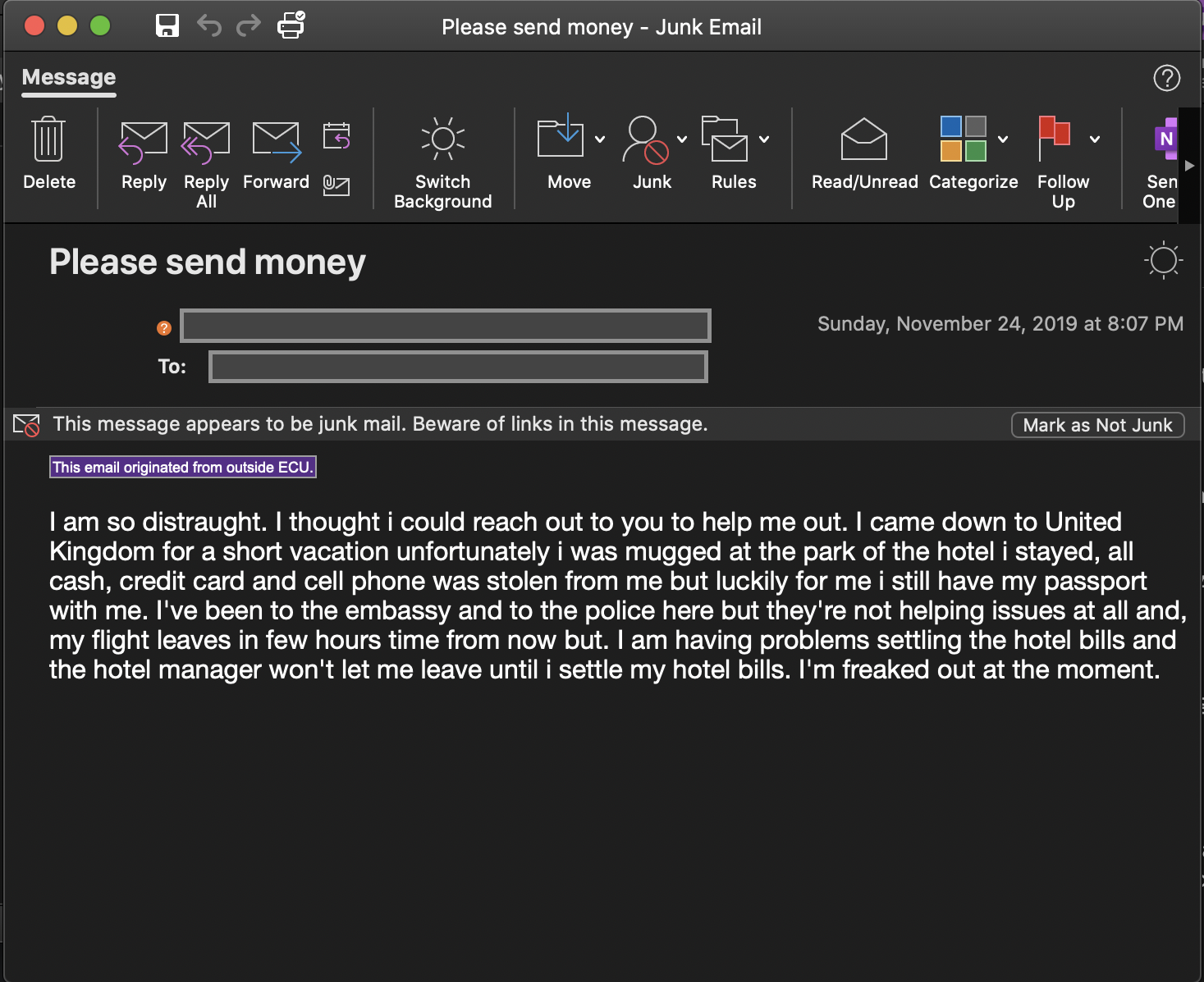}
    \caption{Email with unchanged text is flagged as spam.}\label{fig:Microsoft email: junk}
\end{subfigure}
~
\begin{subfigure}[t]{0.48\textwidth} 
    \centering
    \includegraphics[width=\textwidth]{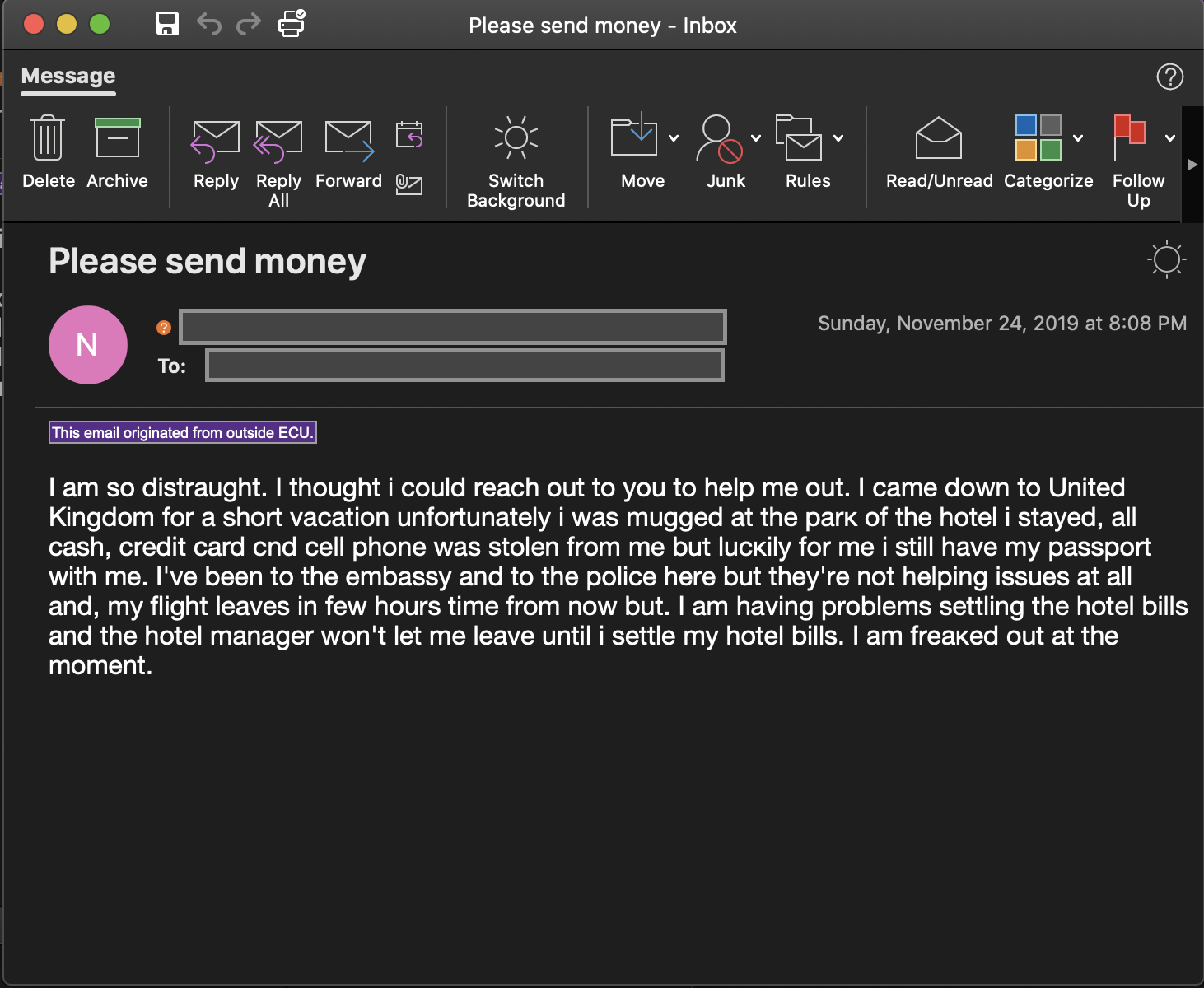}
    \caption{Email with text changed using confusables by-passes the spam filter and is delivered to the inbox.}\label{fig:Microsoft email: inbox}
\end{subfigure}

\caption{Based on the content of this email, the spam filter of the email server flags it as spam (left), whereas by changing some of the characters with their corresponding confusables tricks the spam filter into marking this message as safe (right).}\label{fig:Microsoft email}
\end{figure*}

\section{Conclusion and Future Work}

Visual spoofing has been around for many years however it is still a major problem with web security. The rapid growth of web use of Unicode characters has brought in new and more advanced forms of visual attacks. The key cause of the attacks presented in this work, is the presence in Unicode space of several identical characters that can be used by the attackers to confuse users.

In this paper we presented a method that can be used to avoid detection by a spam filter. With this approach, a sender can replace a limited number of letters from the Latin alphabet with letters from other alphabet(s) that look alike, and in doing so it tricks the spam filter to produce more errors. We evaluated this method using publicly available copies of spam and ham emails, namely from Enron1 data set, with four machine learning algorithms: decision trees, random forests, na\"ive Bayes, and support vector machine. Our experiments indicate that using a classifier trained on data using Latin alphabet, to classify a message with a combination of Latin and Cyrillic letters leads to much lower classification accuracy compared to the same classifier used with a message with Latin characters only.

Moreover, we tested this method with a Microsoft Business email. We first sent an email containing a lot of keywords frequently encountered in spam emails, and this email was flagged as spam. Then we sent the same email, with some of the characters replaced by their ``visually equivalent'' characters from Cyrillic alphabet, and this email was delivered to the Inbox. This suggests that this method can currently bypass existing spam filters.

Although we evaluated this method in the context of spam filtering, this has implications for other text communication and documents, as described in Section~\ref{Results and Discussion}. Examples include avoiding plagiarism detection by automated software, eluding detection when sending malicious messages with instant messaging applications, and a range of other applications that use natural language processing for automatic analysis of text documents.

In future work we plan to evaluate this approach with characters from multiple alphabets. In addition, we would like to investigate the impact of this method with other applications used for text communication.

\bibliographystyle{acm}
\bibliography{visual_spoofing}

\end{document}